\documentclass[conference]{IEEEtran}

\usepackage[hidelinks]{hyperref}
\usepackage{graphicx}
\usepackage{graphics}
\usepackage{stackengine}
\usepackage{algpseudocode}
\usepackage{listings}
\usepackage{amsmath}
\usepackage{balance}
\usepackage{color}
\usepackage{enumitem}
\usepackage{listings}
\usepackage{adjustbox}
\usepackage[caption=false]{subfig}
\usepackage{xspace}
\usepackage{color}
\usepackage{multirow}
\usepackage{listings}
\usepackage{comment}
\usepackage{hhline}
\usepackage{tabularx}
\usepackage{tcolorbox}

\usepackage{caption}
\usepackage{multicol}

\usepackage{fancyvrb}

\usepackage{setspace}
\usepackage[linesnumbered,vlined,boxed,ruled,nofillcomment,noend]{algorithm2e}
\SetKw{Continue}{continue}
\SetKw{Break}{break}
\usepackage{algpseudocode}
\newlength{\oldtextfloatsep}

\usepackage{graphicx}
\usepackage{seqsplit}
\usepackage{cancel}

\newcommand{\Ignore}[1]{}

\newcommand{\ie}{i.e.,\xspace}
\newcommand{\eg}{e.g.,\xspace}
\newcommand{\etc}{etc.\xspace}
\newcommand{\tech}{\textsc{ARUS}\xspace}
\newcommand{\techc}{\textsc{ARUS}\textsubscript{C}\xspace}
\newcommand{\evalprojnum}{128\xspace}
\newcommand{\evalusd}{280\xspace}
\newcommand{\evalusdr}{276\xspace}
\newcommand{\evaluso}{1,529\xspace}
\newcommand{\percreso}{66.8\%\xspace}
\newcommand{\percresd}{98.6\%\xspace}
\newcommand{\mergednum}{83\xspace}
\newcommand{\evalprojnumwithus}{40\xspace}

\hypersetup{
colorlinks=false,
bookmarksnumbered=true,
}

\newcommand\opt[1]{}
\newcommand\find[1]{}

\newcommand{\ls}[1]
   {\dimen0=\fontdimen6\the\font 
    \lineskip=#1\dimen0
    \advance\lineskip.5\fontdimen5\the\font
    \advance\lineskip-\dimen0
    \lineskiplimit=.9\lineskip
    \baselineskip=\lineskip
    \advance\baselineskip\dimen0
    \normallineskip\lineskip
    \normallineskiplimit\lineskiplimit
    \normalbaselineskip\baselineskip
    \ignorespaces
   }

\newenvironment{smalldescription}{
   \setlength{\topsep}{0pt}
   \setlength{\partopsep}{0pt}
   \setlength{\parskip}{0pt}
   \begin{description}
   \setlength{\leftmargin}{.2in}
   \setlength{\parsep}{0pt}
   \setlength{\parskip}{0pt}
   \setlength{\itemsep}{0pt}}{\end{description}}

\newcounter{observation}
\mathchardef\mhyphen="2D

\newcommand*\Reactivatenumber[1]{%
  \lst@AddToHook{OnNewLine}{%
   \let\thelstnumber\origthelstnumber%
   \setcounter{lstnumber}{\numexpr#1-1\relax}%
  }%
}

\newcommand{\ssymbol}[1]{^{\@fnsymbol{#1}}}

\let\oldnl\nl
\newcommand{\nonl}{\renewcommand{\nl}{\let\nl\oldnl}}%

\makeatletter
\renewcommand{\@IEEEsectpunct}{\ \,}
\makeatother

\begin{document}

\title{Automatically Removing Unnecessary Stubbings\\from Test Suites}

\author{\IEEEauthorblockN{Mengzhen Li, Mattia Fazzini}
\IEEEauthorblockA{\textit{University of Minnesota, MN, USA}; li001618@umn.edu, mfazzini@umn.edu}
}

\maketitle

\begin{abstract}
Most modern software systems are characterized by a high number of components whose interactions can affect and complicate testing activities. During testing, developers can account for the interactions by isolating the code under test using test doubles and stubbings. During the evolution of a test suite, stubbings might become unnecessary, and developers should remove unnecessary stubbings, as their definitions can introduce unreliable test results in future versions of the test suite. Unfortunately, removing unnecessary stubbings is still a manual task that can be complex and time-consuming.

To help developers in this task, we propose \tech, a technique to automatically remove unnecessary stubbings from test suites. Given a software project and its test suite, the technique executes the tests to identify unnecessary stubbings and then removes them using different approaches based on the characteristics of the stubbings. We performed an empirical evaluation based on \evalprojnum Java projects that use Mockito for stubbing and contain \evalusd stubbing definitions that lead to \evaluso unnecessary stubbings. Overall, our technique provides a solution for \evalusdr of the definitions (\percresd resolution rate), \tech' time cost is negligible, and, on average, the technique's changes introduce a limited increase in code complexity. We submitted \tech' changes to the projects through pull requests and \mergednum resolutions are already merged.

\end{abstract}
 
\section{Introduction}

Nowadays, we use software for many of our daily activities, such as shopping, banking, and social networking. Due to the importance that software has in our daily lives, software must be tested to gain confidence it behaves correctly. Because software is often characterized by a multitude of interacting components (\eg software units, libraries, web services, \etc), during testing, it can be desirable for developers to isolate the code under test from the components that the code depends on. To that end, developers can use \textit{test doubles}\footnote{Although test doubles are often informally called \textit{mocks}, in this work we use test doubles as it is a more formal and general term.}~\cite{2023_fowler_testdoubles, 2007_meszaros_xunit}.

In class-based, object-oriented programming, test doubles are objects that mimic the structure of other objects but offer alternative implementations that developers can fully control for testing purposes. Test doubles are specialized for the purpose of a test and can be used to (i) simply fill-in parameters that are meaningless for a specific test, (ii) return hard-coded values when their methods are invoked, (iii) verify interactions with other classes, and (iv) provide partially working implementations that are more efficient than the ones provided by the actual objects they are replacing. A key and often used feature offered by test doubles is the ability to define objects that return hard-coded values when their methods are invoked~\cite{2022_icse_fazzini_use,2019_ese_spadini_mock,2017_msr_spadini_to-mock}. In this case, the methods are called \textit{stubbed methods} or \textit{stubbings}.

Developers can define test doubles and their stubbings using test mocking frameworks\footnote{Although these frameworks are informally called test mocking frameworks, developers actually use the frameworks to create test doubles.} (\eg Mockito~\cite{2023_mockito}, EasyMock~\cite{2023_easymock}, PowerMock~\cite{2023_powermock}, \etc). With these frameworks, developers can define test doubles and stubbings that can be used by a single test or reused by multiple tests.

Because software projects evolve continuously, developers need to modify and add tests to their test suites to ensure the software continues to behave as expected. In this process, certain stubbed methods may become \textit{unnecessary stubbings}---methods that are stubbed but never executed during a specific test execution. This situation might be an oversight of the developer, the artifact of copy-paste, or the effect of not fully understanding the test or code. Unnecessary stubbings can be seen as an instance of the general fixture test smell~\cite{2001_zxp_van_refactoring} and should be removed to keep the test code clean, reliable, and maintainable~\cite{2023_unusedstubbings}. To remove unnecessary stubbings, developers need to deal with the definitions creating the unnecessary stubbings. Although developers wish to remove unnecessary stubbings~\cite{2023_github_downlords-faf-client, 2023_resolveunusedstubbings}, they can find it complex or time-consuming to do so~\cite{2023_github_downlords-faf-client}.

Recent work proposed techniques~\cite{2020_ase_zhu_mocksniffer,2021_fse_wang_an,zhu2023stubcoder} and performed studies~\cite{2014_icqs_mostafa,2017_icst_trautsch_are,2017_msr_spadini_to-mock,2019_ese_spadini_mock,2020_icsme_pereira_assessing,2022_icse_fazzini_use} to help developers create and maintain test doubles. However, to the best of our knowledge, no prior work focused on automatically removing unnecessary stubbings.
Wang et al.~\cite{2021_fse_wang_an} presented an approach for automatically refactoring test doubles that are built through inheritance (often referred to as mock classes) with test doubles that are built using Mockito. Pereira and Hora~\cite{2020_icsme_pereira_assessing} characterized the use of mock classes in Java programs. Other studies analyzed the use of test doubles in Java~\cite{2014_icqs_mostafa,2017_msr_spadini_to-mock,2019_ese_spadini_mock}, Android~\cite{2022_icse_fazzini_use}, and Python~\cite{2017_icst_trautsch_are}. The studies identified commonly used test mocking frameworks and categorized the classes that tend to be replaced with test doubles. Fazzini et at.~\cite{2022_icse_fazzini_use} also identified that unnecessary stubbings are common.

In this paper, we present \tech, a technique for automatically removing unnecessary stubbings from test suites. The technique takes as inputs the software under test and its test suite. \tech produces as outputs an updated test suite, where unnecessary stubbings have been suitably removed, and a report detailing the changes in the test suite. The technique operates in three phases. First, \tech executes the test suite and, for each test, collects information about its stubbing definitions, stubbing invocations, and unnecessary stubbings. Second, the technique converts the ``dynamic'' information collected through test execution into ``static''
test code information that can be used to handle the definitions leading to unnecessary stubbings.
Third, \tech uses different resolution approaches to remove unnecessary stubbings having different characteristics and produces a report describing the changes.

We implemented \tech in a prototype tool that supports Java programs with Java test suites using Mockito for stubbing. (We focused on tests written in Java, as it is one of the most popular programming languages~\cite{2023_github_top_languages}, and Mockito, as it is the most widely used test mocking framework for Java-based software~\cite{2014_icqs_mostafa,2022_icse_fazzini_use}). We used the prototype tool to perform an empirical evaluation of \tech. The evaluation is based on \evalprojnum Java projects that contain \evalusd stubbing definitions leading to \evaluso unnecessary stubbings. The evaluation investigates whether the technique is able to successfully remove unnecessary stubbings, reports on the number of resolutions accepted by the developers of the projects under analysis, assesses the time cost of \tech, and studies the variation in code complexity of the updated test suites. The technique was able to resolve \evalusdr stubbing definitions leading to unnecessary stubbings (\percresd resolution rate) and developers already merged \mergednum of the resolutions (37 were not and the remaining are awaiting review). We also identified that \tech' time cost is negligible and, on average, the technique introduces only a limited increase in code complexity.

In summary, this paper makes the following contributions:
\begin{itemize}[noitemsep,topsep=2pt,parsep=0pt,partopsep=0pt,left=10pt]
\item An automated technique called \tech that automatically removes unnecessary stubbing from test suites.
\item An empirical evaluation that provides initial evidence of the effectiveness and efficiency of our technique.
\item An implementation of the technique that is publicly available, together with the evaluation infrastructure.~\cite{2023_arus_artifact}.
\end{itemize}

\section{Background}
\label{sec:back}

In this section, we provide background on testing with test doubles, introduce key terminology, and present a motivating example that includes a test with unnecessary stubbings.

\newsavebox\componentundertest
\begin{lrbox}{\componentundertest}
\begin{lstlisting}[escapechar=|,language=Java,basicstyle=\scriptsize\ttfamily,numbers=left,numbersep=2pt, numberstyle=\scriptsize\color{black},frame=None,xleftmargin=30pt,xrightmargin=0pt,aboveskip=5pt,belowskip=5pt]
pulic class ChangesSinceLastUnstableBuildMacroTest {
 ...
 @Test
 public void testShouldReverseOrderOfChanges() ... {|\label{example:motiv:testReverseOrderedContentChanges_start}| 
  content.reverse = true;
  AbstractBuild failBld = createBuild(Result.FAILURE,|\label{example:motiv:invk1_start}|
   41, "Changes for a failed build.");|\label{example:motiv:invk1_end}|
  AbstractBuild currBld = createBuild(Result.SUCCESS,|\label{example:motiv:invk2_start}|
   42, "Changes for a successful build.");|\label{example:motiv:invk2_end}|
  when(currBld.getPreviousBuild()).thenReturn(failBld);|\label{example:motiv:s1_end}|
  when(failBld.getNextBuild()).thenReturn(currBld);|\label{example:motiv:us1_end}|
  String contentStr = content.evaluate(currBld, lis, 
   ChangesSinceLastUnstableBuildMacro.MACRO_NAME);
  assertEquals(..., contentStr);|\label{example:motiv:assert1_end}|
 }|\label{example:motiv:testReverseOrderedContentChanges_end}|
 private AbstractBuild createBuild(Result result, |\label{example:motiv:createbuild_start}|
  int buildNumber, String message) {
  AbstractBuild build = mock(AbstractBuild.class);|\label{example:motiv:td}|
  when(build.getResult()).thenReturn(result);|\label{example:motiv:us2}|
  ChangeLogSet changes1 = createChangeLog(message);
  when(build.getChangeSet()).thenReturn(changes1);|\label{example:motiv:us3}|
  when(build.getChangeSets()).thenReturn(|\label{example:motiv:s4_start}|
   Collections.singletonList(changes1));|\label{example:motiv:s4_end}|
  when(build.getNumber()).thenReturn(buildNumber);|\label{example:motiv:s5}|
  return build;
 }|\label{example:motiv:createbuild_end}|
}
\end{lstlisting}
\end{lrbox}

In class-based, object-oriented programming, a test suite (TS) for the software under test (SUT) includes tests

that exercise components of the SUT. These components are also known as the components under test (CUTs). Tests are generally divided into multiple test classes.
Each test class can also contain a \textit{setup} method, which holds the code that sets the pre-conditions for correctly executing the tests. Tests are generally designed to have four parts, which execute in sequence. The four parts are: \textit{setup}, \textit{exercise}, \textit{verify}, and \textit{teardown}~\cite{2007_meszaros_xunit}. The setup part sets the pre-conditions for correctly executing a specific test. The exercise part interacts with the $\mathit{CUT}$. The verify part checks that the expected outcome was obtained. The teardown part puts the state of the SUT back into what it was before executing the test. Test doubles and stubbings are generally created in a setup method or the setup part of a test.

In Java, JUnit~\cite{2023_junit} and Mockito~\cite{2023_mockito} are the most popular frameworks for creating tests and test doubles~\cite{2014_icqs_mostafa,2022_icse_fazzini_use}. Using JUnit, developers can define tests by marking test class methods using the {\small\texttt{@Test}} annotation. Developers can define setup methods using the {\small\texttt{@Before}} annotation\footnote{Different JUnit versions have slightly different constructs for marking tests and setup methods, but all recent versions of JUnit offer those capabilities~\cite{2023_junit}.}. With Mockito, developers can define test doubles using the {\small\texttt{mock(...)}} method from the framework's API~\cite{2023_mockito_api}. This method creates an object of the type passed as the parameter, and the object's methods have empty implementations. Developers can define stubbings on the test double using the {\small\texttt{when(...).thenReturn(...)}} ``pattern'' from the framework's API. In the pattern, the {\small\texttt{when}} method takes as input the method call whose behavior is being defined. The {\small\texttt{thenReturn}} method takes as input the value that the test double returns when the method call provided to the {\small\texttt{when}} method is triggered during test execution. Mockito also offers other methods (\eg~{\small\texttt{spy(...)}}) and patterns (\eg~{\small\texttt{doReturn(...).when(...).method(...)}}) to create test doubles and stubbings.

When a test defines a stubbing, the stubbing can be either used or unnecessary for the test. When the code location defining a stubbing is shared across tests (\eg when it appears in a setup method) or it is used multiple times by a test (\eg when the definition appears in a \textit{helper} method), the location leads to multiple stubbing definitions and each of the definitions can be either used or unnecessary for a test. When we want to highlight that an unnecessary stubbing is from a specific test execution, we use the term \textit{unnecessary stubbing occurrence}. We use the term \textit{stubbing definition occurrence} (or \textit{stubbing occurrence} in short) to refer to a specific stubbing definition appearing in a specific test execution.

To motivate our work, we now discuss a real-world example of a test class containing unnecessary stubbings. The example is taken from the test suite of the \textsc{Token} project~\cite{2023_github_token}, which is a Jenkins~\cite{jenkins} plugin with 39 contributors. Figure~\ref{fig:motiv} reports a test ({\small\texttt{testShouldReverseOrderOfChanges}}) and a helper method ({\small\texttt{createBuild}}) from the test~class~{\small\texttt{ChangesSinceLastUnstableBuildMacroTest}}\footnote{In the figure, we abbreviated the test name, shorten some variable names, and omitted some parts (\eg additional tests) due to space limitations.}. The test uses the helper method twice (lines~\ref{example:motiv:invk1_start}$\mhyphen$\ref{example:motiv:invk1_end} and~\ref{example:motiv:invk2_start}$\mhyphen$\ref{example:motiv:invk2_end}). At each invocation, the helper method creates a test double of type {\small\texttt{AbstractBuild}} (line~\ref{example:motiv:td}), defines four stubbings on the test double (lines~\ref{example:motiv:us2},~\ref{example:motiv:us3},~\ref{example:motiv:s4_start}$\mhyphen$\ref{example:motiv:s4_end}, and~\ref{example:motiv:s5}), and returns the test double. The stubbing defined at line~\ref{example:motiv:us3} through the first invocation of the helper method and the stubbings defined at lines~\ref{example:motiv:us2} and~\ref{example:motiv:us3} through the second invocation of the method are unnecessary for the test. The helper method is also invoked by 11 other tests and for all the tests the stubbing defined at line~\ref{example:motiv:us3} is unnecessary. Furthermore, the stubbing defined at line~\ref{example:motiv:us1_end} is also unnecessary for the test. This example shows that it is possible to have a code location that defines the same stubbing multiple times (for the same test or multiple tests) but some definitions are necessary and others are not. Additionally, the example also shows that there might be stubbings that are always unnecessary and those can be defined in helper methods or in the test itself. Removing unnecessary stubbings requires a careful analysis of the tests and their executions.

\begin{figure}[!t]
\begin{minipage}[!t]{0.99\linewidth}
\centering
{
\usebox\componentundertest
}
\end{minipage}
\vspace{-8pt}
\caption{Example of a test with unnecessary stubbings.}
\label{fig:motiv}
\vspace{-18pt}
\end{figure}

\section{Technique}
\label{sec:technique}

\begin{figure}[!t]
  \centerline{\includegraphics[width=1.02\columnwidth]{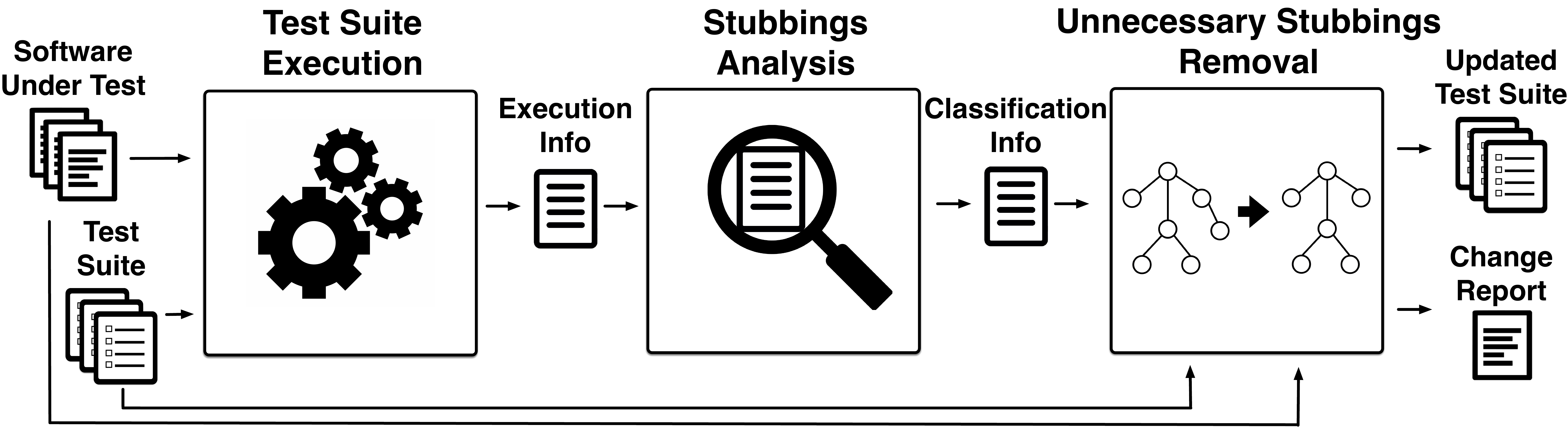}}
  \caption{High-level overview of \tech.}
  \label{fig:technique}
  \vspace{-22pt}
\end{figure}

In this section, we present \tech, our technique for automatically removing unnecessary stubbings from a test suite. Figure~\ref{fig:technique} provides an overview of \tech and shows its three main phases. Given the SUT and its TS\footnote{\tech requires all tests in the TS to pass and it assumes that unnecessary stubbings are unnecessary as all tests pass.}, the \emph{test suite execution} phase executes the TS on the SUT and, for each test, collects information about stubbing definitions, stubbing invocations, and unnecessary stubbings. \tech logs that information in the \textit{execution info} file. The \emph{stubbings analysis} phase processes collected information and classifies unnecessary stubbings into three types. \tech classifies unnecessary stubbings into the three types by looking at the information provided by the execution of all the tests. Finally, the \emph{unnecessary stubbings removal} phase leverages the classification information to suitably change the code of the test suite and remove unnecessary stubbings. \tech' final output is an \textit{updated test suite} (UTS) and a report (R) detailing the changes appearing in the UTS.

\subsection{Test Suite Execution}
\label{sec:tse}

This phase collects information about stubbings and unnecessary stubbings in the TS. Specifically, this phase executes the TS on the SUT and, for each test, collects information about the stubbings in the test. At a high level, this phase collects three main pieces of information: the stubbings defined by the test, the test code location where the stubbings are defined, and whether the stubbings are unnecessary stubbings. \tech stores collected information in the execution info file. To collect the desired information, \tech uses an instrumented version of the test mocking framework used by the TS. Although it would be possible to collect the same information by instrumenting both the SUT and the TS, we believe that using an instrumented version of the test mocking framework limits the amount of instrumentation needed for the task and has a lower impact on the execution time of the TS. Figure~\ref{fig:executioninfo} reports part of the information contained in the execution info file for the execution of the {\small\texttt{testShouldReverseOrderOfChanges}} test, which we presented in the motivating example of Section~\ref{sec:back}). The rest of this section details the high level concepts and reports examples based on the content of Figure~\ref{fig:executioninfo}. (The line numbers in Figure~\ref{fig:executioninfo} do not match the line numbers in Figure~\ref{fig:motiv}, as Figure~\ref{fig:motiv} is a simplified version of the actual code in the \textsc{Token} project.)

For each test execution, \tech collects the name of the test class (\eg\xspace{\small\texttt{ChangesSinceLast UnstableBuildMacroTest}}), the name of the test (\eg\xspace{\small\texttt{testShouldReverseOrderOfChanges}}), the start and the end of when the test executes, the stubbings that are defined, the invocation of stubbed methods, and the unnecessary stubbings. Unnecessary stubbings are computed at the end of the test execution based on defined stubbings and the stubbed methods that are never invoked. \tech collects all the above information for each test as stubbings might be used by some tests but not by others.

For each stubbing definition, \tech assigns an identifier to the stubbing (\eg\xspace{\small\texttt{AbstractBuild\#getResult\#102}}) and saves the class name of the method being stubbed (\eg\xspace{\small\texttt{AbstractBuild}}), the name of the method being stubbed (\eg\xspace{\small\texttt{getResult}}), the test code location where the stubbing is defined (\eg line 344 in {\small\texttt{ChangesSinceLastUnstableBuildMacroTest.java}}), and the stack trace at the time the stubbing is being defined. \tech collects stack trace information to differentiate between usages of multiple definitions of the same stubbing (as in the example of Section~\ref{sec:back} when a helper method that defines a stubbing is called multiple times within the same test). \tech logs stubbing definitions by intercepting when the test mocking framework creates those stubbings (\eg when the {\small\texttt{when(...).thenReturn(...)}} ``pattern'' is invoked).

\begin{figure}[t]
\centering
\begin{minipage}[t]{.99\linewidth}
\begin{Verbatim}[fontsize=\scriptsize, commandchars=\\\{\}]
\(\textbf{test_start}\)
\(\textbf{test_method_execution_start}\)
\(\textbf{test_method_class:}\)...ChangesSinceLastUnstableBuildMacroTest
\(\textbf{test_method_name:}\)testShouldReverseOrderOfChanges
...
\(\textbf{stub_creation_info_start}\)
\(\textbf{stubbing_id:}\)...AbstractBuild#getResult#102
\(\textbf{stubbed_method_class:}\)...AbstractBuild
\(\textbf{stubbed_method_name:}\)getResult
\(\textbf{stubbing_location:}\)...createBuild(...MacroTest.java:344)
\(\textbf{stack:}\)...MacroTest.java;...;createBuild;344#
...MacroTest.java;...;testShouldReverseOrderOfChanges;85#...
\(\textbf{stub_creation_info_end}\)
...
\(\textbf{test_method_execution_end}\)
\(\textbf{unnecessary_stubbing_info_start}\)
\(\textbf{stubbing_id:}\)...AbstractBuild#getResult#102
\(\textbf{stubbed_method_class:}\)...AbstractBuild
\(\textbf{stubbed_method_name:}\)getResult
\(\textbf{stubbing_location:}\)...createBuild(...MacroTest.java:344)
\(\textbf{unnecessary_stubbing_info_end}\)
\(\textbf{test_end}\)
\end{Verbatim}
\vspace{-5pt}
\caption{Portion of an execution info file.}
\label{fig:executioninfo}
\end{minipage}
\vspace{-20pt}
\end{figure}

For each stubbing invocation, \tech logs the identifier of the stubbing, the class name of the method being invoked, the name of the method being invoked, the code location where the method is being invoked, and the test code location where the stubbing was defined. \tech identifies invocations by intercepting the stubbed method call made on the test double.

\tech determines unnecessary stubbings by identifying stubbing definitions that do not have any corresponding stubbing invocation. For each unnecessary stubbing, \tech logs the identifier of the stubbing (\eg\xspace{\small\texttt{AbstractBuild\#getResult\#102}}) the class name of the method being stubbed (\eg\xspace{\small\texttt{AbstractBuild}}), the name of the method being stubbed (\eg\xspace{\small\texttt{getResult}}), and the test code location where the stubbing was defined (\eg line 344 in {\small\texttt{ChangesSinceLastUnstableBuildMacroTest.java}}).

After executing the TS on the SUT, \tech passes the execution info file to the next phase of the technique.

\subsection{Stubbings Analysis}

This phase classifies unnecessary stubbings into different types so that \tech can apply different resolution approaches based on the nature of the stubbings. At a high level, this phase first merges stubbing occurrences from different tests by the code location of the stubbing definitions and then classifies unnecessary stubbings into three types: \textit{totally-unnecessary} (TU), \textit{used-unnecessary-setup} (UUS), and \textit{used-unnecessary-helper} (UUH) stubbings. A TU stubbing is a stubbing that is defined but never used by any of the tests. TU stubbings can be defined in a test (\eg the stubbing defined at line~\ref{example:motiv:us1_end} in Figure~\ref{fig:motiv}), in a setup method, or in an helper method (\eg the stubbing defined at line~\ref{example:motiv:us3} in Figure~\ref{fig:motiv}). UUS and UUH stubbings are stubbings that sometimes are used but sometimes are unnecessary. UUS stubbings identify stubbings whose definition is created through a setup method (either directly in the method or indirectly by calling other methods). UUH stubbings are stubbings defined in helper methods, which, in \tech' context, are methods called directly or indirectly by a test and never called by a setup method. We propose and use these three categories as it is possible to resolve the three types using different resolution approaches. We provide more details on how \tech groups and classifies unnecessary stubbings with the help of Algorithm~\ref{alg:c}.

Algorithm~\ref{alg:c} takes as inputs the test suite ($\mathit{TS}$) and the execution info file ($\mathit{eif}$). The algorithm's output is a set of classified unnecessary stubbings ($\mathit{cusSet}$ in the algorithm and \textit{classification info} in Figure~\ref{fig:technique}). A classified unnecessary stubbing ($\mathit{cus}$) is a static abstraction of the dynamic (\ie test execution) information contained in the $\mathit{eif}$, that is, a $\mathit{cus}$ groups together the information of stubbing definitions that occur at a specific test code location and at least one of the occurrences leads to an unnecessary stubbing. Considering the test in Figure~\ref{fig:motiv} and the associated execution info file in Figure~\ref{fig:executioninfo}, the stubbing definition at line~\ref{example:motiv:us2} occurs two times and one of those is unnecessary. The two occurrences would be grouped together into a single $\mathit{cus}$ by \tech.

\setlength{\textfloatsep}{0.001cm}
\setlength{\floatsep}{0.001cm}

\SetInd{4.5pt}{4.5pt}
\begin{algorithm}[t]
\begin{scriptsize}
    \DontPrintSemicolon
    \SetKwInOut{Input}{Input}\SetKwInOut{Output}{Output}
    \caption{Unnecessary Stubbings Classification}
    \label{alg:c}
    \Input{
        $\mathit{TS}$: the test suite; $\mathit{eif}$: the execution info file\\ 
    }
    \Output{
        $\mathit{cusSet}$: set of classified unnecessary stubbings\\
    }
    \Begin{
        $\mathit{cusSet}~$=$~\emptyset$\;\label{alg:c:emptyInfoList}
        $\mathit{gsSet}~$= \textsc{Group-Stubbings($\mathit{eif}$)}\;\label{alg:c:groupus}
        \ForEach{\textup{$\mathit{gs} \in \mathit{gsSet}$}}{\label{alg:c:identification-begin}
            $\mathit{tusdList}$ = $\mathit{gs}$.\textsc{Get-Tests-Of-USD()}\;\label{alg:c:identification-prop-start}
            $\mathit{tisdList}$ = $\mathit{gs}$.\textsc{Get-Tests-Of-ISD()}\;\label{alg:c:identification-prop-end}
            \If{\textup{$\mathit{usdList} == []$ }}{
                \textit{// no unnecessary stubbings in the group}\;\label{alg:c:continue-start}
                \Continue\;\label{alg:c:continue-end}
            }
            \If{\textup{$\mathit{tisdList} == []$ }}{
                \textit{// totally-unnecessary case}\;\label{alg:c:tu-begin}
                $\mathit{cus}$= \textsc{Create-CUS($\mathit{gs}$,TU)}\;
                $\mathit{cusSet}$.\textsc{Conditional-Add($\mathit{TS}$,$\mathit{cus}$)}\;\label{alg:c:tu-end}
           }
           \Else{\label{alg:partially-unused-begin}
                \If{\textsc{Through-Setup($\mathit{gs}$)}}{\label{alg:c:uus-begin}
                    \textit{// used-unnecessary-setup case}\;
                    $\mathit{cus}$= \textsc{Create-CUS($\mathit{gs}$,UUS)}\;\label{alg:c:uus-crated}
                }
                \Else{\label{alg:c:uuh-begin}
                    \textit{// used-unnecessary-helper case}\;
                    $\mathit{cus}$= \textsc{Create-CUS($\mathit{gs}$,UUH)}\;\label{alg:c:uuh-created}
                }\label{alg:c:uuh-end}
                $\mathit{cusSet}$.\textsc{Conditional-Add($\mathit{TS}$,$\mathit{cus}$)}\;\label{alg:c:uu-added}
           }\label{alg:partially-unused-end}   
        }\label{alg:c:identification-end}
        \Return $\mathit{cusSet}$\;
    }
\end{scriptsize}
\end{algorithm}

A $\mathit{cus}$ contains the type of unnecessary stubbing (TU, UUS, or UUH), the list of stubbing definitions that led to the stubbing being unnecessary ($\mathit{usdList}$), the list of tests that led to $\mathit{usdList}$ ($\mathit{tusdList}$), the (possibly empty) list of stubbing definitions that share the same code location as the definitions in $\mathit{usdList}$ but whose stubbings were invoked in some tests ($\mathit{isdList}$), and the list of tests that led to $\mathit{isdList}$ ($\mathit{tisdList}$). The code location of the stubbing definitions in $\mathit{usdList}$ uniquely identifies a $\mathit{cus}$. Although the stubbing definitions in $\mathit{usdList}$ share the same code location, \tech considers all of them as their stack traces might be different (\eg invocations of {\small\texttt{createBuild}} in Figure~\ref{fig:motiv}). Furthermore, $\mathit{usdList}$ and $\mathit{tusdList}$ always have at least one element because a $\mathit{cus}$ represents at least one unnecessary stubbing.

The algorithm starts by grouping stubbing definition occurrences appearing in the $\mathit{eif}$ (\textsc{Group-Stubbings}) based on their code location. The result of this operation is a set of grouped stubbings ($\mathit{gsSet}$). A group of stubbings ($\mathit{gs}$) is an abstraction that contains the same information as a $\mathit{cus}$ except that it does not have a label for the categorization of the unnecessary stubbing(s) it represents and that the $\mathit{usdList}$ might be empty (as a $\mathit{gs}$ might identify stubbing definitions that are always used). After creating $\mathit{gsSet}$, the algorithm enters its main loop
and categorizes the unnecessary stubbings. Each iteration focuses on one group of stubbings at a time.

At the beginning of each iteration, the algorithm retrieves some properties of the $\mathit{gs}$. Specifically, the algorithm retrieves the unnecessary stubbing definitions ($\mathit{usdList}$) and the test information ($\mathit{tusdList}$ and $\mathit{tisdList}$) associated with the $\mathit{gs}$ (lines~\ref{alg:c:identification-prop-start}$\mhyphen$\ref{alg:c:identification-prop-end}). If the $\mathit{gs}$ does not have any definition that was unnecessary, then the algorithm moves to the next group of stubbings (lines~\ref{alg:c:continue-start}$\mhyphen$\ref{alg:c:continue-end}), as no change is needed. If no test in the TS invokes the stubbing (lines~\ref{alg:c:tu-begin}$\mhyphen$\ref{alg:c:tu-end}), the algorithm classifies the stubbing as totally-unnecessary, creates a $\mathit{cus}$ using $\mathit{gs}$ and the classification, and adds the $\mathit{cus}$ to the result set ($\mathit{cusSet}$) only after checking some additional properties of the stubbing (\textsc{Conditional-Add}). With the \textsc{Conditional-Add} function, the algorithm checks whether the stubbing is defined within a parameterized test or a loop by analyzing the abstract syntax tree (AST) of the test code in the TS. If that is the case, the algorithm does not add the $\mathit{cus}$ to the $\mathit{cusSet}$ as more information might be needed to suitably modify the test code related to the $\mathit{cus}$. In the case of parameterized tests, information about test parameters might be needed. For loops, information about the loop index might be needed. Although it is possible to collect this information with additional test code instrumentation, we decided not to do so as the instrumentation might be significant and, therefore, affect test execution. We leave the investigation of these cases as future work. The other classifications of the algorithm (UUS and UUH) also use the check before adding the $\mathit{cus}$ to the result set ($\mathit{cusSet}$).

If the $\mathit{gs}$ has definitions created through a setup method (\textsc{Through-Setup}), the algorithm categorizes the stubbing as a used-unnecessary-setup stubbing (line~\ref{alg:c:uus-crated}). Otherwise, the algorithm identifies the stubbing as a used-unnecessary-helper stubbing (line~\ref{alg:c:uuh-created}). In both scenarios, the result of the classification ($\mathit{cus}$) is added to the ($\mathit{cusSet}$). To give an example, when the algorithm considers the stubbing definition at line~\ref{example:motiv:us2} of Figure~\ref{fig:motiv}, the {\small\texttt{testShouldReverseOrderOfChanges}} test is both in the $\mathit{tisdList}$ and the $\mathit{tusdList}$ as the stubbing is both used and unnecessary in the test, and the algorithm classifies the stubbing as of type UUH.

After grouping and classifying the unnecessary stubbings, \tech updates the test suite based on the content of the $\mathit{cusSet}$. We describe this part of the technique next.

\subsection{Unnecessary Stubbings Removal}

This phase edits the code of the TS to remove unnecessary stubbings. At a high level, the phase identifies the test files containing stubbing definitions leading to unnecessary stubbings and changes the files using different resolution approaches based on how the unnecessary stubbings were classified by the previous phase of the technique. We now detail how this phase operates with the help of Algorithm~\ref{alg:r}.

Algorithm~\ref{alg:r} requires three main inputs: the software under test ($\mathit{SUT}$), the test suite ($\mathit{TS}$), and the classification of the unnecessary stubbings ($\mathit{cusSet}$). The algorithm also uses an optional flag ($\mathit{auusr}$) that developers can use to control removal of UUS stubbings. Specifically, we give developers the opportunity to avoid removal of UUS stubbings as their removal requires introducing new test classes with modified setup methods (more details in the rest of this section) and this operation can increase the code complexity of the test suite. The output of the algorithm is an updated test suite ($\mathit{UTS}$) and a report ($\mathit{R}$) describing the changes in the UTS.

\begin{algorithm}[!t]
\begin{scriptsize}
    \DontPrintSemicolon
    \SetKwInOut{Input}{Input}\SetKwInOut{Output}{Output}
    \caption{Unnecessary Stubbings Removal}
    \label{alg:r}
      \Input{
        $\mathit{SUT}$: the software under test; $\mathit{TS}$: the test suite; $\mathit{cusSet}$: set of classified unnecessary stubbings; $\mathit{auusr}$: boolean to avoid removal of used-unnecessary-setup stubbings ($\mathit{false}$ by default)\\
      }
       \Output{
        $\mathit{UTS}$: updated test suite; $\mathit{R}$: report describing the changes in UTS\\
       }
       \Begin{
       $\mathit{UTS}~$=$~$\textsc{Copy-TS($\mathit{TS}$)}\;\label{alg:r:copy}
       $\mathit{R}~$=$~\emptyset$\;\label{alg:r:empty-report}
       $\mathit{cusSetMap}~$=$~$\textsc{Group-By-File($\mathit{cusSet}$,$\mathit{UTS}$)}\;
        \ForEach{\textup{$\mathit{testFilePath} \in \mathit{cusSetMap}$.\textsc{Keys()}}}{\label{alg:r:main-loop-start}
            $\mathit{ast}~$=$~$\textsc{Parse-File($\mathit{testFilePath}$,$\mathit{UTS}$)}\;\label{alg:r:parse-file}
            \ForEach{\textup{$\mathit{cus} \in \mathit{cusSetMap}$.\textsc{Get($\mathit{testFilePath}$)}}}{\label{alg:r:inner-loop-start}
                \If{\textup{$\mathit{cus}$.\textsc{Get-Type()} == $~\mathit{TU}$}}{
                    \textit{// process totally-unnecessary stubbings}\;
                    \textsc{Code-Removal($\mathit{ast}$,$\mathit{cus}$)}\;\label{alg:r:tu}
                }
                \ElseIf{\textup{$\mathit{cus}$.\textsc{Get-Type()} == $~\mathit{UUH}$}}{
                    \textit{// process used-unnecessary-helper stubbings}\;
                    \textsc{Method-Duplication-And-Removal($\mathit{ast}$,$\mathit{cus}$)}\;\label{alg:r:uuh}
                }
                \ElseIf{\textup{(($\mathit{cus}$.\textsc{Get-Type()} == $~\mathit{UUS}$)~$\land$~($\neg\mathit{auusr}$))}}{
                    \textit{// process used-unnecessary-setup stubbings}\;
                    \textsc{Class-Duplication-And-Removal($\mathit{ast}$,$\mathit{cus}$)}\;\label{alg:r:uus}
                }
            }\label{alg:r:inner-loop-end}
            \textsc{Store-File($\mathit{ast}$,$\mathit{testFilePath}$,$\mathit{UTS}$)}\;\label{alg:r:store-file}     
        }\label{alg:r:main-loop-end}
        $\mathit{R}~$=$~$\textsc{Create-Report($\mathit{TS}$,$\mathit{UTS}$)}\;\label{alg:r:report}
        \Return $\mathit{UTS}$,$\mathit{R}$\; 
    }   
\end{scriptsize}
\end{algorithm}

The algorithm begins by initializing the $\mathit{UTS}$ to be the same as the $\mathit{TS}$ and $\mathit{R}$ to be empty (lines~\ref{alg:r:copy}$\mhyphen$\ref{alg:r:empty-report}). After this initial step, the algorithm groups (\textsc{Group-By-File}) classified unnecessary stubbings from the same test file into a map ($\mathit{cusSetMap}$) so that the algorithm can process all the unnecessary stubbings in one file at once. The absolute path of the file is the identifier for the groups in the map. At this point, the algorithm enters its main loop and starts processing the files that require change.

In the first step of the iteration, the algorithm parses the file (\textsc{Parse-File}) into its corresponding abstract syntax tree ($\mathit{ast}$). The algorithm makes changes to the file through the $\mathit{ast}$. At this point, the algorithm processes all the classified unnecessary stubbings ($\mathit{cus}$) in the file.

If the current $\mathit{cus}$ represents a totally-unnecessary stubbing, the algorithm uses a \textit{code-removal} approach to resolve the unnecessary stubbing (line~\ref{alg:r:tu}). Specifically, the algorithm navigates the $\mathit{ast}$ and removes the stubbing definition associated with the unnecessary stubbing from the code. This solution is suitable because the stubbing is never invoked.

When the current $\mathit{cus}$ represents a used-unnecessary-helper or used-unnecessary-setup stubbing, the algorithm uses a \textit{code-duplication-and-removal}
approach to resolve the unnecessary stubbing. If the stubbing definition associated with the $\mathit{cus}$ appears in a helper method (UUH stubbing), the algorithm (i) duplicates the method, (ii) removes the stubbing definition from the duplicated method, and (iii) replaces calls to the original method with calls to the duplicated method only for those calls that led to the definition of an unnecessary stubbing.

If the stubbing definition is in a setup method (UUS stubbing), the algorithm (i) creates a new test class, (ii) copies the setup method in the new class, (iii) moves the tests having the unnecessary stubbing into the new class, and (iv) removes the stubbing definition from the setup method. In the case of UUS stubbings, the definition could also be in a method whose invocation originates from the setup method. In this case, the algorithm (i) creates a new test class, (ii) copies the setup method in the new class, (iii) moves the tests having the unnecessary stubbing into the new class, (iv) copies the helper method into the new class, (v) duplicates the helper method, (vi) removes the stubbing definition from the duplicated method, and (vii) replaces calls to the original method with calls to the duplicated method only for those calls that led to the definition of an unnecessary stubbing. When necessary, the algorithm also copies other fields or methods in the new test class based on the content of the methods involved in the resolution. We believe that this solution is appropriate as we group tests in different test classes based on the setup/preconditions that the tests need. However, considering that this operation can copy multiple code entities in new test classes, and that can impact test code complexity, \tech also allows developers to avoid removal of UUS stubbings.

\tech' code duplication approach can handle stubbing definitions in methods at any call-chain distance from the test or the setup method and minimizes the number of new test classes. To perform the code duplication task, the algorithm leverages the information contained in the stack traces that are associated with the stubbing definitions in the $\mathit{cus}$.

To give an example, the algorithm resolves the UUH stubbing at line~\ref{example:motiv:us2} in Figure~\ref{fig:motiv} by duplicating the {\small\texttt{create Build}} helper method, removing the stubbing {\small\texttt{when(build. getResult()).thenReturn(result)}} from the duplicated method, and calling the duplicated method on lines~\ref{example:motiv:invk2_start}-\ref{example:motiv:invk2_end}.

After processing all the classified unnecessary stubbings in a file, the algorithm saves the updated $\mathit{ast}$ into the corresponding file of the $\mathit{UTS}$. When the algorithm finishes processing all the files that require change, it prepares $\mathit{R}$ by describing the changes in the $\mathit{UTS}$ and returns the $\mathit{UTS}$ and $\mathit{R}$ to the developer. The $\mathit{UTS}$ and $\mathit{R}$ are \tech' final outputs.

\section{Implementation}

We implemented \tech in a tool that is publicly available~\cite{2023_arus_artifact} and consists of three main modules. Each module corresponds to a phase in \tech. The \textit{test execution} module extends Mockito and uses Java to configure the Maven~\cite{2023_maven} build file ({\small\texttt{pom.xml}}) of the SUT to run its test suite with our extended version of Mockito. Although Mockito can be configured to report unnecessary stubbings, the information provided by Mockito is not enough to automatically resolve them. We extended Mockito by adding 806 lines of code. The \textit{stub analysis} and the \textit{unnecessary stubbings removal} modules are implemented through 10,272 lines of code. The modules use Java Parser~\cite{2023_javaparser} to process the ASTs of the test files.

\section{Evaluation}
This section discusses our empirical evaluation. To assess the efficacy and the efficiency of \tech, we investigated the following research questions:

\begin{itemize}[leftmargin=3pt,labelsep=2pt]
\item \textbf{RQ1}: \textit{Can \tech remove unnecessary stubbings \\{\color{white}--------}from test suites?}
\item \textbf{RQ2}: \textit{What is the time cost of running \tech?}
\item \textbf{RQ3}: \textit{How does \tech affects test code complexity?}
\item \textbf{RQ4}: \textit{What are the developers' reactions to \tech' changes?}
\end{itemize}

While presenting the result for \tech in its default configuration, we also discuss the performance of \tech without UUS stubbing removal. We use \techc to refer to this second version of the technique.

\subsection{Benchmarks}

We used a set of \evalprojnum real-world Java projects to evaluate \tech. We identified this benchmark set based on a dataset of 147,991 Java projects made available by related work~\cite{2019_arxiv_loriot_styler,2020_arxiv_soto-valero_coverage-based,2021_fse_soto-valero_a}. The dataset contains GitHub~\cite{2023_github} projects whose primary language is Java and that have at least five stars. The projects were retrieved from GitHub in June 2020. To the best of our knowledge, the dataset is the largest, most recent, and readily available dataset of Java projects from GitHub.

From the dataset, we first selected all projects that use Maven as their build system as the implementation of \tech supports Maven to automatically build the projects and execute their tests. To identify those projects, we cloned all the projects in the dataset and, for each project, checked whether it contained a single {\small\texttt{pom.xml}} file. We looked for projects with a single {\small\texttt{pom.xml}} file as that provides a good filter for working with the primary module of a project. This step left us with 33,418 projects (20,794 projects had multiple {\small\texttt{pom.xml}} files and 93,779 projects had zero).

\begin{table*}[!t]
  \setlength\tabcolsep{6pt}
  	\caption{\small{Projects with unnecessary stubbings.
  	For each project:
  	\textit{ID} = identifier;
  	\textit{Name} = name;
    \textit{Commit ID} = version analyzed; 
    \textit{LOC$_{sc}$} = \# of code lines in the source code;
    \textit{LOC$_{tc}$} = \# of code lines in the test code;
    \textit{Tests} = \# of tests;
  	\textit{USD$_{b}$} = \# of stubbings definitions leading to unnecessary stubbings before running \tech;
  	\textit{USD$_{a}$} = \# of stubbings definitions leading to unnecessary stubbings after running \tech;
    \textit{USO$_{b}$} = \# of unnecessary stubbing occurrences before running \tech;
  	\textit{USO$_{a}$} = \# of unnecessary stubbing occurrences after running \tech;
    \textit{TU} =  \# of totally-unnecessary stubbings;
    \textit{UUH} =  \# of used-unnecessary-helper stubbings;
    \textit{UUS} =  \# of used-unnecessary-setup stubbings.
    }}
  	\vspace{-10pt}
  	\label{tab:benchmarks}
  	\begin{footnotesize}
  	\begin{center}
  	\begin{tabular}[h]{|l|l|c|r|r|r||r|r|r|r||r|r|r|}
    \hhline{------||----||---}
  	\multicolumn{6}{|c||}{\textit{Benchmarks}} &
  	\multicolumn{4}{c||}{\textit{Unnecessary Stubbings}}&
    \multicolumn{3}{c|}{\textit{Types}} \\
  	\hhline{------||----||---}
    \multicolumn{1}{|c}{\textit{ID}} &
    \multicolumn{1}{|c}{\textit{Name}} &
    \multicolumn{1}{|c}{\textit{Commit ID}} &
    \multicolumn{1}{|c}{\textit{LOC$_{sc}$}} &
    \multicolumn{1}{|c}{\textit{LOC$_{tc}$}} &
    \multicolumn{1}{|c||}{\textit{Tests}} &
    \multicolumn{1}{|c}{\textit{USD}$_{b}$} &
    \multicolumn{1}{|c}{\textit{USD}$_{a}$} &
    \multicolumn{1}{|c}{\textit{USO}$_{b}$} &
    \multicolumn{1}{|c||}{\textit{USO}$_{a}$} &
    \multicolumn{1}{c|}{\textit{TU}} &
    \multicolumn{1}{c|}{\textit{UUH}} &
    \multicolumn{1}{c|}{\textit{UUS}} \\
    \hhline{------||----||---}
    \noalign{\vspace{1.5pt}}
   	\hhline{------||----||---}

    \textsc{P01} & \href{https://github.com/allure-framework/allure-bamboo}{\underline{\textsc{Allure}}} & 85a9408c
    & $1086$ & $135$ & $15$ 
    & $2$ & $0$ & $5$ & $0$ 
    & $1$ & $0$ & $1$\\

     \textsc{P02} & \href{https://github.com/jenkinsci/amazon-ecs-plugin}{\underline{\textsc{Amazon-ecs}}} & 44817eda 
    & $2403$ & $231$  & $27$
    & $3$ & $0$ & $3$ & $0$
    & $3$ & $0$  & $0$    \\

     \textsc{P03} & \href{https://github.com/awslabs/amazon-sqs-java-extended-client-lib}{\underline{\textsc{Amazon-sqs}}} & 450d5221
    & $481$ & $417$ & $35$ 
    & $1$ & $0$ & $14$ & $0$  
    & $0$ & $0$  & $1$\\

     \textsc{P04} & \href{https://github.com/jenkinsci/appcenter-plugin}{\underline{\textsc{Appcenter}}} & 986ec689
    & $1810$ & $1301$ & $146$ 
    & $7$ & $0$ & $26$ & $0$  
    & $0$ & $0$  & $7$\\

    \textsc{P05} & \href{https://github.com/awslabs/aws-codepipeline-custom-job-worker}{\underline{\textsc{Aws-codepipeline}}} & 33952495
    & $762$ & $525$ & $45$ 
    & $2$ & $0$ & $16$ & $0$  
    & $0$ & $0$  & $2$\\

    \textsc{P06} & \href{https://github.com/blockchain-jd-com/bftsmart}{\underline{\textsc{Bftsmart}}} & 44c1cb2e
    & $16053$ & $3379$ & $12$ 
    & $8$ & $1$ & $42$ & $2$ 
    & $7$ & $0$ & $0$\\

    \textsc{P07} & \href{https://github.com/apereo/cas-server-security-filter}{\underline{\textsc{CAS}}} & a84c946c
    & $552$ & $357$ & $38$ 
    & $1$ & $0$ & $1$ & $0$  
    & $1$ & $0$ & $0$   \\

    \textsc{P08} & \href{https://github.com/jenkinsci/chucknorris-plugin}{\underline{\textsc{Chucknorris}}} & 2a9dc4b0
    & $159$ & $208$ & $26$ 
    & $4$ & $0$ & $12$ & $0$  
    & $0$ & $0$  & $4$\\

    \textsc{P09} & \href{https://github.com/RentTheRunway/conduit}{\underline{\textsc{Conduit}}} & c6f82f67
    & $1400$ & $1415$ & $79$ 
    & $24$ & $0$ & $112$ & $0$ 
    & $17$ & $0$ & $7$ \\

    \textsc{P10} & \href{https://github.com/jenkinsci/datadog-plugin}{\underline{\textsc{Datadog}}} & 875c82b9
    & $7670$ & $4127$ & $161$ 
    & $17$ & $0$ & $34$ & $0$ 
    & $14$ & $1$ & $2$ \\

   \textsc{P11} & \href{https://github.com/vandeseer/easytable}{\underline{\textsc{Easytable}}} & b3c278a9
    & $1661$ & $1082$ & $57$ 
    & $4$ & $0$ & $8$ & $0$ 
    & $1$ & $3$  & $0$\\

    \textsc{P12} & \href{https://github.com/jenkinsci/github-branch-source-plugin}{\underline{\textsc{Github-branch}}} & e708675a
    & $7056$  & $5924$ & $522$
    & $2$ & $0$ & $3$ & $0$  
    & $2$ & $0$  & $0$  \\

    \textsc{P13} & \href{https://github.com/weswilliams/GivWenZen}{\underline{\textsc{GivWenZen}}} & 455a03aa
    & $1757$ & $1089$ & $57$ 
    & $2$ & $0$ & $2$ & $0$ 
    & $1$ & $0$  & $1$\\

    \textsc{P14} & \href{https://github.com/jenkinsci/google-compute-engine-plugin}{\underline{\textsc{Google-compute}}} & 08e2f706
    & $2166$ & $1279$ & $50$ 
    & $19$ & $0$ & $76$ & $0$ 
    & $7$ & $1$ & $11$ \\

   \textsc{P15} & \href{https://github.com/jenkinsci/google-kubernetes-engine-plugin}{\underline{\textsc{Google-kubernetes}}} & ef890e4a
    & $999$ & $1089$ & $109$ 
    & $8$ & $0$ & $36$ & $0$  
    & $0$ & $7$ & $1$ \\
    
    \textsc{P16} & \href{https://github.com/jenkinsci/google-oauth-plugin}{\underline{\textsc{Google-oauth}}} & 3e03b2cb
    & $1021$ & $1303$ & $89$ 
    & $4$ & $0$ & $4$ & $0$  
    & $4$ & $0$ & $0$\\

    \textsc{P17} & \href{https://github.com/hap-java/HAP-Java}{\underline{\textsc{HAP}}} & f4a9872d
    & $6066$ & $98$ & $12$ 
    & $2$ & $0$  & $7$ & $0$  
    & $0$ & $0$  & $2$\\

    \textsc{P18} & \href{https://github.com/jenkinsci/hashicorp-vault-plugin}{\underline{\textsc{Hashicorp}}} & 182c0fba
    & $2477$ & $2028$ & $110$ 
    & $15$ & $0$ & $144$ & $0$ 
    & $3$ & $10$  & $2$ \\

    \textsc{P19} & \href{https://github.com/jenkinsci/instant-messaging-plugin}{\underline{\textsc{Instant-messaging}}} & 51f23def
    & $2992$ & $674$ & $41$ 
    & $16$ & $2$ & $521$ & $503$ 
    & $10$ & $4$  & $0$\\

   \textsc{P20} & \href{https://github.com/KittehOrg/KittehIRCClientLib}{\underline{\textsc{KittehIRCClientLib}}} & 46b57952
    & $9938$ & $2168$ & $198$ 
    & $17$ & $0$ & $77$ & $0$ 
    & $9$ & $1$  & $7$\\

   \textsc{P21} & \href{https://github.com/ldbc/ldbc_snb_datagen}{\underline{\textsc{LDBC}}} & 0c019a46
    & $4736$ & $488$ & $3$
    & $1$ & $0$ & $1$ & $0$  
    & $1$ & $0$  & $0$  \\

  	\textsc{P22} & \href{https://github.com/matomo-org/matomo-java-tracker}{\underline{\textsc{Matomo}}} & 751823e6
    & $1285$ & $1503$ & $184$ 
    & $8$ & $0$ & $8$ & $0$
    & $8$ & $0$ & $0$   \\ 

    \textsc{P23} & \href{https://github.com/garbagemule/MobArena}{\underline{\textsc{MobArena}}} & 9164b125
    & $13906$ & $2572$ & $293$ 
    & $4$ & $0$ & $8$ & $0$ 
    & $1$ & $2$  & $1$\\
   
   \textsc{P24} & \href{https://github.com/MutabilityDetector/MutabilityDetector}{\underline{\textsc{MutabilityDetector}}} & ac1bc226
    & $6710$ & $3421$ & $371$ 
    & $5$ & $0$ & $5$ & $0$  
    & $5$ & $0$ & $0$  \\

    \textsc{P25} & \href{https://github.com/Juniper/netconf-java}{\underline{\textsc{Netconf}}} & c0fbedac
    & $1570$ & $545$ & $51$ 
    & $2$ & $0$ & $14$ & $0$  
    & $0$ & $0$ & $2$\\

    \textsc{P26} & \href{https://github.com/curityio/oauth-filter-for-java}{\underline{\textsc{Oauth-filter}}} & eb27b214
    & $979$ & $241$ & $11$ 
    & $3$ & $1$ & $5$ & $3$ 
    & $2$ & $0$ & $0$   \\

   \textsc{P27} & \href{https://github.com/Terracotta-OSS/offheap-store}{\underline{\textsc{Offheap}}} & 05cc59ec
    & $11180$ & $7192$ & $125$ 
    & $1$ & $0$ & $1$ & $0$  
    & $1$ & $0$ & $0$  \\

    \textsc{P28} & \href{https://github.com/mdewilde/opml-parser}{\underline{\textsc{OPML}}} & ae6a03d9
    & $904$ & $935$ & $58$ 
    & $1$ & $0$ & $1$ & $0$  
    & $1$ & $0$  & $0$  \\

    \textsc{P29} & \href{https://github.com/cloudspannerecosystem/pgadapter}{\underline{\textsc{Pgadapter}}} & e64d3f0d 
    & $2759$ & $1205$ & $86$ 
    & $12$ & $0$ & $10$ & $0$ 
    & $7$ & $0$  & $5$\\

   \textsc{P30} & \href{https://github.com/komoot/photon}{\underline{\textsc{Photon}}} & 4343b9f3  
    & $2201$ & $1513$ & $112$ 
    & $5$ & $0$ & $22$ & $0$ 
    & $3$ & $2$  & $0$\\

    \textsc{P31} & \href{https://github.com/s-webber/projog}{\underline{\textsc{Projog}}} &  70fea568
    & $9761$ & $9623$ & $1100$ 
    & $1$ & $0$ & $53$ & $0$  
    & $0$ & $0$ & $1$ \\

    \textsc{P32} & \href{https://github.com/jenkinsci/repository-connector-plugin}{\underline{\textsc{Repository-connector}}} & 34fef47d 
    & $1418$ & $577$ & $23$ 
    & $1$ & $0$ & $3$ & $0$  
    & $0$ & $0$  & $1$\\ 

    \textsc{P33} & \href{https://github.com/Invictum/serenity-reportportal-integration}{\underline{\textsc{Serenity}}} & 4c5476f3
    & $861$ & $523$ & $80$ 
    & $2$ & $0$ & $3$ & $0$  
    & $0$ & $0$ & $2$ \\

   \textsc{P34} & \href{https://github.com/apache/sling-org-apache-sling-commons-threads}{\underline{\textsc{Sling}}} & ff2418ae
    & $1209$ & $189$ & $9$ 
    & $1$ & $0$ & $1$ & $0$  
    & $1$ & $0$ & $0$\\

    \textsc{P35} & \href{https://github.com/vaulttec/sonar-auth-oidc}{\underline{\textsc{Sonar-auth}}} & 99d86044
    & $395$ & $727$ & $65$ 
    & $25$ & $0$ & $73$ & $0$ 
    & $15$ & $6$  & $4$\\

   \textsc{P36} & \href{https://github.com/perforce/sonar-scm-perforce}{\underline{\textsc{Sonar-scm}}} & 115cc273
    & $357$ & $104$  & $6$ 
    & $5$ & $0$ & $5$ & $0$  
    & $5$ & $0$  & $0$\\

    \textsc{P37} & \href{https://github.com/jenkinsci/subversion-plugin}{\underline{\textsc{Subversion}}} & dd1693c1
    & $6463$ & $2517$ & $293$ 
    & $2$ & $0$ & $13$ & $0$ 
    & $1$ & $1$   & $0$\\

   \textsc{P38} & \href{https://github.com/dmulloy2/SwornAPI}{\underline{\textsc{SwornAPI}}} & 0e33d2a1
    & $4182$ & $102$ & $7$ 
    & $4$ & $0$ & $20$ & $0$ 
    & $2$ & $2$ & $0$ \\

    \textsc{P39} & \href{https://github.com/jenkinsci/token-macro-plugin}{\underline{\textsc{Token}}}  & 871c6edc
    & $2214$ & $1884$ & $176$ 
    & $37$ & $0$ & $135$ & $0$ 
    & $27$ & $10$  & $0$\\

     \textsc{P40} & \href{https://github.com/jenkinsci/xunit-plugin}{\underline{\textsc{Xunit}}} & bf2a9c19
    & $2078$ & $1246$ & $191$ 
    & $2$ & $0$ & $5$ & $0$  
    & $0$ & $0$ & $2$ \\

   
	\hhline{------||----||---}
    \noalign{\vspace{1.5pt}}
   	\hhline{------||----||---}
    \multicolumn{3}{|r|}{$Total$}&
    \multicolumn{1}{r}{$143677$}&
    \multicolumn{1}{|r}{$65936$}&
    \multicolumn{1}{|r||}{$5073$}&
    \multicolumn{1}{|r}{$280$}&
    \multicolumn{1}{|r}{$4$}&
    \multicolumn{1}{|r}{$1529$}&
    \multicolumn{1}{|r||}{$508$}&
    \multicolumn{1}{r|}{$160$}&
    \multicolumn{1}{r}{$50$}&
    \multicolumn{1}{|r|}{$66$}\\

    \hhline{------||----||---}
  	\end{tabular}
  	\end{center}
  	\end{footnotesize}
  	\vspace{-20pt}
\end{table*}

We then identified projects that potentially had JUnit tests by looking for those projects that declared JUnit-related dependencies. We identified this information using the dependency tree provided by Maven~\cite{2023_maven_dependency-tree}. This step left us with 14,568 projects. We checked for projects that potentially created test doubles with Mockito in a similar way, and that step left us with 1,562 projects. Amongst these remaining projects, 904 used a version of Mockito compatible with our technique. Mockito 2.3.0 or greater can be configured so that it does not affect the test outcome regardless of whether the tests have unnecessary stubbings. Across the remaining projects, 540 have tests that execute and pass over three runs. We ran the tests three times to filter out projects that had flaky tests. From the remaining 540 projects, we selected those using JUnit 4 and Mockito 3, and that left us with our final set of \evalprojnum projects. We focused on JUnit 4 and Mockito 3 as those are the versions supported by the implementation of our technique. Additionally, JUnit 4 was the version used by the highest number of projects (379), and Mockito 3 was the version under active maintenance used by the highest number of projects (\evalprojnum) when we identified the dataset.

\subsection{Methodology}

To answer the research questions, we ran \tech and \techc on the benchmarks using a workstation with 128GB of memory, one Intel Core i9-9900K 3.60GHz processor, and running Ubuntu 16.04. To answer RQ1, RQ2, and RQ3, we used the most recent version of the benchmark projects when we finished collecting the dataset (January 2022). To validate the results of RQ1, we manually inspected the results and ran the updated test suites on the benchmarks. All the tests in all the updated test suites passed. To measure the time cost of \tech and \techc (RQ2), we ran the experiments three times and presented execution time statistics from those runs. To measure how \tech and \techc affect code complexity (RQ3), we compare the test code before and after \tech' and \techc' changes using two metrics: percentage difference in cognitive complexity~\cite{munoz2020empirical} and percentage difference in the lines of code (LOC). We used cognitive complexity as it positively correlates with comprehension time~\cite{munoz2020empirical}. To measure cognitive complexity, we used Genese Cpx~\cite{genese-cpx}. To measure LOC, we use CK~\cite{aniche-ck}. The percentage difference computation is based on all the test files in the test suites. Finally, to determine the developers' reactions to \tech' changes, we submitted pull requests containing \tech' changes to the most recent version of the projects in our benchmark set and collected how many of \tech' changes were merged, how many were not, and how many are awaiting review.

\subsection{Results}
\label{sec:results}

\subsubsection*{\textbf{RQ1}: Can \tech remove unnecessary stubbings from test suites?} Out of the \evalprojnum projects in our benchmark dataset, \evalprojnumwithus projects had at least one unnecessary stubbing, which corresponds to $31.3\%$ of the projects in the dataset.

Table~\ref{tab:benchmarks} reports relevant characteristics for the projects having unnecessary stubbings.
In total, the projects have more than 60 thousand lines of test code and five thousand tests.

Overall, \tech provided a solution for \percresd of the definitions that led to unnecessary stubbings, which removed 1021 unnecessary stubbings from the projects. This result corresponds to \percreso of all the unnecessary stubbings in all test executions. If we remove an outlier case (\textsc{P19}), which has 503 unnecessary stubbings in a loop, the percentage increases to 99.5\%.
The columns under the \textit{Unnecessary Stubbings} and \textit{Types} report details on the results.
Columns \textit{USD$_{b}$}, \textit{USD$_{a}$}, \textit{USO$_{b}$}, \textit{USO$_{a}$}, \textit{TU}, \textit{UUH} and \textit{UUS} report the number of stubbing definitions leading to unnecessary stubbings before running \tech, the number of stubbing definitions leading to unnecessary stubbings after running \tech, the number of unnecessary stubbing occurrences before running \tech, the number of unnecessary stubbing occurrences after running \tech, the number of unnecessary stubbings classified as totally-unnecessary, used-unnecessary-helper, and used-unnecessary-setup, respectively. Note that columns \textit{USD$_{b}$}/\textit{USD$_{a}$} and \textit{USO$_{b}$}/\textit{USO$_{a}$} can differ as the same stubbing definition might lead to multiple unnecessary stubbings. The results presented in Table~\ref{tab:benchmarks} were validated through manual inspections and by running the updated test suites, which had all tests passing.

Out of the 280 stubbings resolved by \tech 160 are of type TU, 50 of type UUH, and 66 of type UUS. \tech was not able to provide a solution for only four stubbing definitions (column \textit{USD$_{a}$}). One definition was part of a parameterized test and three definitions were inside loops. Although this result is good as the number of definitions not handled by \tech (by design) is extremely low, two of the definitions inside project \textsc{P19}~led to 503 unresolved unnecessary stubbings. We investigated the nature of those definitions and they used the loop index. Although this case can be seen as an outlier based on the benchmarks in our dataset, the resolution of those definitions might be an interesting direction for future work.

\techc was able to resolve 210 stubbings (160 of type TU and 50 of type UUH). Although \techc is able to resolve a lower number of stubbings as compared to \tech, we believe that this second version of the technique is still effective and can be helpful for handling unnecessary stubbings.

\begin{tcolorbox}[colback=white, boxsep=0pt, left=4pt, right=4pt, before skip=4pt,after skip=8pt]
\textbf{RQ1 answer:} \tech was able to provide a solution for \percresd of the unnecessary stubbing definitions, which removed 1021 unnecessary stubbings from the projects. Based on the results, we can say that \tech is effective in removing unnecessary stubbings.
\end{tcolorbox}

\subsubsection*{\textbf{RQ2}: What is the time cost of running \tech?} To evaluate the time cost of \tech, we measured the difference in the execution time of the tests when using the original version of Mockito versus our customized version, the time required to run \tech' analyses (stubbings analysis and unnecessary stubbings removal phases), and the difference in the test execution time between the original test suite and the updated one. This evaluation is based on the \evalprojnumwithus projects with unnecessary stubbings. We report all the measurements in seconds (s).

The execution time of the test suites with our customized version of Mockito was on average 0.99s (max=25.99s, min=-45.28s, sd=8.8s) slower with respect to the execution of the test suites with the original version of Mockito. \tech' analyses took 0.533s on average (max=3.962s, min=0.001s, sd=0.862s). \tech' changes led to updated test suites that are 2.49s slower with respect to the original test suites (max=80.27s, min=-49.95s, sd=16.07s). The results indicate that the cost of running \tech is low as running the test suites with our version of Mockito adds just a few seconds to their execution, \tech' analyses take only a few seconds, and the execution of the updated test suites is a few seconds longer compared to the original ones. Furthermore, \tech does not necessarily need to be run with all the test suite runs.

Considering \techc, the analyses took 0.529s on average, and \techc' updated test suites are 0.35s slower with respect to the original test suites. Based on these results, we can say that the time cost of \tech and \techc are similar.

\begin{tcolorbox}[colback=white, boxsep=0pt, left=4pt, right=4pt, before skip=4pt,after skip=8pt]
\textbf{RQ2 answer:} Based on the low execution time of \tech' analyses and the low increase in the execution time of the updated test suites, we can say that the time cost associated with \tech is low.
\end{tcolorbox}

\subsubsection*{\textbf{RQ3}: How does \tech affect test code complexity?} 
\begin{figure}[!t]
    \includegraphics[width=1.0\linewidth]{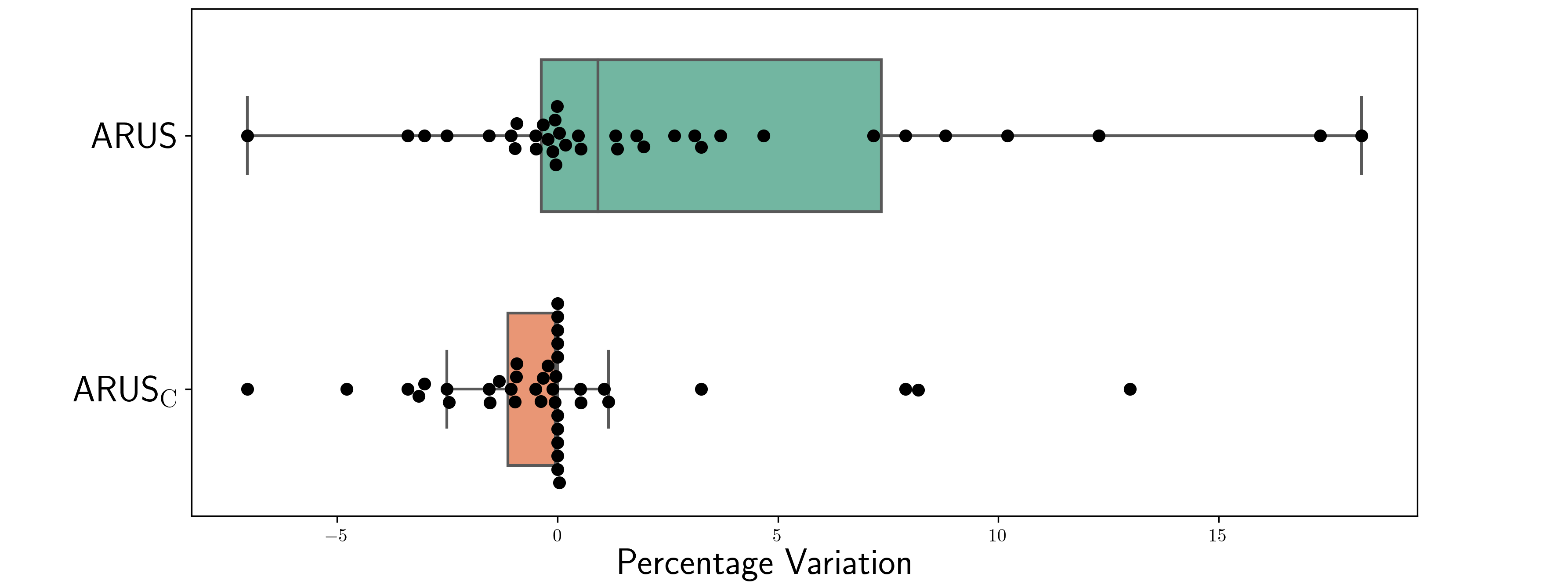}
    \vspace{-17pt}
    \caption{Percentage variation of cognitive complexity.}
    \label{fig:cogperc} 
\end{figure}

\begin{figure}[!t]
    \includegraphics[width=1.0\linewidth]{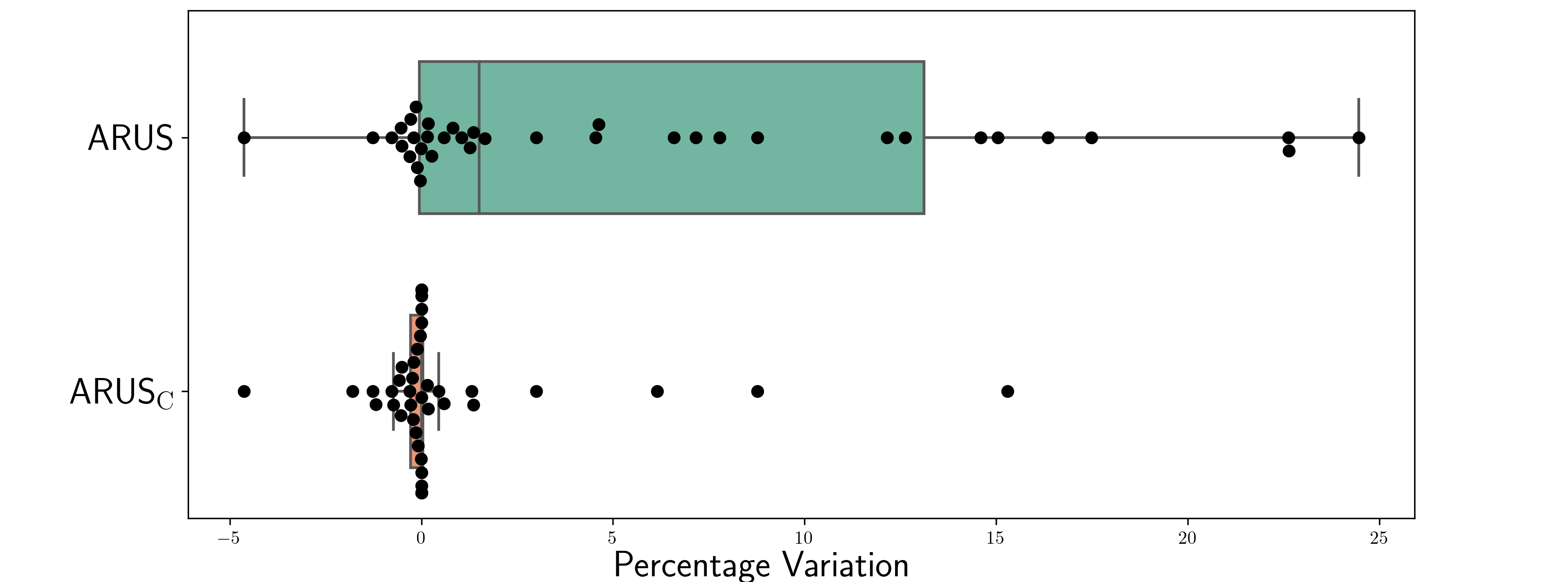}
    \vspace{-17pt}
    \caption{Percentage variation of LOC.}
    \label{fig:locperc} 
    \vspace{-2pt}
\end{figure}

We answer RQ3 with the help of Figure~\ref{fig:cogperc} and~\ref{fig:locperc}. Figure~\ref{fig:cogperc} represents the percentage increase of cognitive complexity for the updated test suites. Figure~\ref{fig:locperc} represents the percentage increase of the LOC for the updated test suites. The figures present results for both \tech and \techc; outliers for \tech are not reported (four in Figure~\ref{fig:cogperc} and three in Figure~\ref{fig:locperc}).


Although in some cases \tech actually reduces code complexity (projects having only TU stubbings) the box plots show that \tech generally introduces an increase in code complexity. This result was expected as the resolution for UUH and USS stubbings, which account for $41.4\%$ of the cases, introduces new code in the test suites.

Although the majority of the projects have an increase in code complexity that we believe to be limited (24 below $5\%$ of both cognitive complexity and LOC increase), eight of the projects had a percentage increase of cognitive complexity and LOC higher than $10\%$. We manually analyzed the changes in those eight projects and identified that all of them had at least one case of a UUS, which required introducing at least one new class (with relevant code) in the test suite. The code complexity results associated with \techc highlight the effect of UUS removal even further. In fact, comparing the results between \tech and \techc from Figure~\ref{fig:cogperc} and~\ref{fig:locperc} it is possible to see that \techc has an overall lower impact on code complexity. These results together with the results from RQ1 highlight a tradeoff between effectiveness and impact on code complexity in resolving unnecessary stubbings. \tech could be used by developers whose priority is on removing unnecessary stubbings while \techc could be used by developers that still interested in removing as many unnecessary stubbings as possible but also prioritize impact on code complexity.

\begin{tcolorbox}[colback=white, boxsep=0pt, left=4pt, right=4pt, before skip=4pt,after skip=8pt]
\textbf{RQ3 answer:} Our results show that \tech generally introduces an increase in code complexity. However, the majority of the projects have an increase in code complexity that we believe to be limited. Furthermore, \tech and \techc can be used to navigate the tradeoff between removal effectiveness and impact on code complexity.
\end{tcolorbox}

\subsubsection*{\textbf{RQ4}: What are the developers' reactions to \tech' changes?} To answer RQ4, we submitted pull requests (PRs) containing \tech' changes to the latest commit of the projects we analyzed. In this process, we submitted PRs to 33 projects (instead of 40 as in Table~\ref{tab:benchmarks}) as two projects were archived, one failed to compile, one changed the build system, one changed test dependencies, one requires Docker support~\cite{docker}, and one had flaky tests. In total, we submitted PRs for 227 unnecessary stubbing definitions. This set does not have two definitions that are in Table~\ref{tab:benchmarks} as they were removed from the corresponding project (P14) and has 13 additional definitions as they were introduced in the projects (P19 and P30).

Overall, developers merged resolutions that removed \mergednum unnecessary stubbing definitions (66 TU, 16 UUH, and one UUS), did not merge 37 (11 TU, 10 UUH, and 16 UUS), and others are awaiting review (54 TU, 21 UUH, and 32 UUS). Although some resolutions were not merged, a good percentage ($36.6\%$) has already been merged. Among the cases that were not merged, we received feedback for one UUS resolution and the developer mentioned that did not want to duplicate code to resolve the unnecessary stubbing. This feedback is in contrast with the merged UUH and UUS resolutions. We believe that \tech and \techc can satisfy different developers' preferences. We also received positive feedback from developers. One developer (from a project that accepted 6 UUH resolutions) mentioned ``\textit{Thanks for this and other PRs, simpler cleaner code is much appreciated [...]}''. Another developer said ``\textit{Good catch, thanks!}''. One more developer expressed interest in \tech: ``\textit{Any place we can read more about this tool, maybe even play with it? [...]}''.

\begin{tcolorbox}[colback=white, boxsep=0pt, left=4pt, right=4pt, before skip=4pt,after skip=8pt]
\textbf{RQ4 answer:} Although a good number of resolutions are awaiting review, developers already merged resolutions that removed \mergednum unnecessary stubbings and only some were not accepted. Based on these preliminary results and the informal feedback we received, we believe that developers had a positive reaction to \tech.
\end{tcolorbox}

\subsection{Discussion}

The results show that \tech is effective in removing unnecessary stubbings.  However, the results also show that some cases are not handled by \tech. These cases are unnecessary stubbings whose definition appears in parametrized tests or loops. These cases could be handled by instrumenting the test cases to observe how stubbing definitions use (directly or indirectly) values of test parameters or loop indexes. Loops might generate a high number of unnecessary stubbings, as shown in the case of project \textsc{P19}. Although loops are not desirable in tests, as they are an instance of the conditional test logic test smell~\cite{2019_cascon_peruma_on}, it might be interesting to investigate resolution approaches for loops and parametrized tests.

While working on \tech, we identified that UUH and UUS stubbings could be alternatively categorized as \textit{used-unnecessary-within-test} (UUWT) and \textit{used-unnecessary-across-tests} (UUAT) stubbings. A UUWT stubbing is characterized by a stubbing definition that creates multiple stubbings in an individual test execution and some of the stubbings are used and others are unnecessary. This is possible when a test method invokes a helper method multiple times. A UUAT stubbing is characterized by a stubbing definition that creates multiple stubbings across different tests and in some tests the stubbings are used while in others the stubbings are unnecessary. Considering the 116 UUS and UUH stubbings in our evaluation, 19 stubbings could be categorized as UUWT and 97 as UUAT. UUAT could be resolved by creating a conditional that executes the stubbing definition based on the name of the test. This solution would have a lower impact on the increase in code complexity of updated test suites. However, the solution would introduce instances of the conditional test logic test smell and bind the test logic to test names. We believe that both traits of the solution are not desirable. UUH and UUS stubbings could also be resolved by relocating the stubbings (and related statements) from helper/setup methods into the tests that need them if that offers a better code complexity as opposed to duplicating methods/classes. Although this solution, when applicable, might limit the code complexity increase, the solution would also spread the logic of helper/setup methods across the tests and developers might not want that to happen as they intentionally created methods to contain the stubbing logic. Future work could investigate whether our sentiment on alternative solutions is shared across practitioners by performing studies with developers.

When we submitted PRs, we noticed that new unnecessary stubbings (13) were introduced in the projects. Future work could study the evolution of unnecessary stubings and provide suggestions on the best time to run \tech.

We checked whether \tech affected the original semantics of the test suites by running the updated test suites on the SUTs and checking that the outcome of the tests did not change. This step provided a validation of our results. The step could also be part of the technique. Specifically, the updated test suite could be run after each change made by \tech. This operation would increase the technique's cost, but the cost could be mitigated through test selection approaches~\cite{engstrom2010systematic,kazmi2017effective}. Future work could investigate the tradeoff provided by different solutions for validating updated test suites.

\section{Limitations and Threats to Validity}

The main limitation of \tech is that it does not handle unnecessary stubbings in parameterized tests or loops. However, our evaluation shows that there is a low number of such cases. In the evaluation, we check for flakiness by running tests three times. Although this might not be enough to detect all the flaky tests, our manual inspection of the results gave us confidence that the results were not affected by flakiness.

As it is the case for most empirical evaluations, there are both external and construct threats to validity associated with the results we presented. In terms of external validity, our results might not generalize to other projects or test suites.
To mitigate this threat, we identified our benchmarks from a large dataset of Java projects having different characteristics. Given the complexity of the projects and the test suites we considered, we believe that \tech should also be applicable to other projects and test suites. Furthermore, we checked \tech' changes in a single testing environment and results might not apply to different testing environments. To mitigate this threat, we inspected the changes manually and we believe that they should not be affected by the testing environment. Additionally, some of the pull requests that we submitted went through a continuous integration pipeline including testing and all tests also passed in those cases. In terms of construct validity, there might be errors in the implementation of our technique. To mitigate this threat, we extensively inspected our code and the results of the evaluation manually.

\section{Related Work}

\noindent\textbf{Studies on Test Doubles.} Test doubles are widely used in software testing to simulate external dependencies, isolate units under test, and facilitate unit testing~\cite{2002_thomas_mock_objects,2006_taeksu_mock_tdd,2022_icse_fazzini_use}. Test doubles have been extensively studied across various programming languages and platforms, including Java~\cite{2014_icqs_mostafa,2019_spadini_mock_java,2019_ese_spadini_mock}, Python~\cite{2017_icst_trautsch_are}, C++~\cite{2022_google_mock_cpp}, and Android~\cite{2022_icse_fazzini_use}. Mostafa and Wang~\cite{2014_icqs_mostafa} identified that Mockito is the most commonly used test mocking framework in Java projects, used in over 70\% of projects that use a test mocking framework. Spadini et al.~\cite{2017_msr_spadini_to-mock,2019_ese_spadini_mock} found that test doubles in Java simplify test structure, prevent contamination of domain code with testing infrastructure, lead to stronger tests and better organization of code, can reduce the cost of writing stubbing code, and make unit testing more effective and efficient. Fazzini et al.~\cite{2022_icse_fazzini_use} identified that test doubles in Android can have issues and unnecessary stubbings are a common problem. Our work proposed a technique for resolving unnecessary stubbings and, through the evaluation, found that unnecessary stubbings appear frequently also in Java projects.

\noindent\textbf{Techniques for Test Doubles.} Related work also proposed techniques for creating and maintaining test doubles. Tillmann and Schulte~\cite{2006_tillmann_mock_gneration} devised an approach that leverages symbolic execution to generate mocks with behavior. Salva and Blot~\cite{2020_salva_model_learninig} proposed a technique that uses a model-based appraoch to generate mocks. Deepika et al.~\cite{2022_tiwari_mimicking} defined an approach that generates tests with mocks from execution data of production code. Zhu et al.~\cite{2020_ase_zhu_mocksniffer} proposed a technique for recommending mocking decisions. Wang et al.~\cite{wang2022jmocker} devised an approach to refactor test doubles built through inheritance with test doubles based on Mockito. Zhu et al.~\cite{zhu2023stubcoder} presented \textsc{StubCoder} a method employing an evolutionary algorithm to automate the synthesis and refinement of stub code for test doubles. \textsc{StubCoder} focuses on making tests pass and, therefore, is not tailored to resolve unnecessary stubbings. \tech focuses on resolving issues in test doubles and, to the best of our knowledge, is the first automated approach to resolve unnecessary stubbings in test suites. Our technique like others in related work~\cite{2022_tiwari_mimicking,2020_ase_zhu_mocksniffer,wang2022jmocker} is based on Mockito.

\noindent\textbf{Test Smells.} Test smells are signs of poorly written test code that negatively affect the quality of test suites, production software, and program comprehension during maintenance activities~\cite{2018_spadini_ts_code_quality,2019_garousi_know_ts_,2015_bavota_ts_harmful}. Researchers have proposed various approaches to detect and address test smells, including metrics-based, rule-based, heuristic-based, static analysis, machine learning, and deep learning approaches~\cite{2007_van_metric_detect_ts,2013_greiler_static_analysis_detect_ts,2010-moha_decor,2019_ananta_detect_ts_pl,2012_maiga_antipattern,2005_marinescu_measurement_quality}. Lambiase et al.~\cite{2020_lambiase_draft} presented a technique that detects and refactors three types of test smells: general fixture, eager test, and lack of cohesion of test methods. Van Deursen et al.~\cite{2001_zxp_van_refactoring} describe 11 test smells and propose refactorings to improve test code for XP practitioners. Our work focuses on automatically removing unnecessary stubbings, which can be seen as a particular type of the general fixture test smell.

\noindent\textbf{Unnecessary Code.} Unnecessary code appears frequently in software systems and increases maintenance costs, making code comprehension difficult~\cite{2012_eder_ucmatter,2020_romano_investigation_into_dead_code}. Tempero~\cite{2008_tempro_unused_design} conducted an empirical study on unused design decisions in open-source Java applications, emphasizing the need for improved tool support to track the unused designs' usage.
Romano and Scanniello~\cite{2018_romano_rta} presented an approach relying on Rapid Type Analysis to detect the dead method smell in Java desktop applications. Romano et al.~\cite{2016_romano_dum} introduced an approach to identify unreachable methods using a graph-based representation of Java software.
Jiang et al.~\cite{2016_jiang_jred} presented a static analysis approach to trim unused methods and classes from both Java application code and the Java Runtime Environment. Heo et al.~\cite{2018_heo_reinforcement_learning} devised a system that uses reinforcement learning to efficiently reduce programs by removing redundant code. Our work removes unnecessary stubbings, which are unnecessary code in test executions.

\section{Conclusion}
In this paper, we presented \tech, an automated technique for removing unnecessary stubbings from test suites. The empirical evaluation of \tech showed that the technique is effective in removing unnecessary stubbings, its cost is negligible, the updated test suites generally have a limited increase in code complexity, and a good number of the resolutions are already merged into the test suites of the projects analyzed.

In terms of future work, we plan to work with developers and investigate their preference on resolution strategies. Future work could also investigate the cost of different solutions for validating \tech' changes. Finally, future work could also investigate approaches to resolve unnecessary stubbing definitions appearing in parameterized tests and loops.

\section*{Acknowledgments}

We thank Mats Heimdahl for several fruitful discussions.
This work is partly supported by funding from Facebook.

\balance

\bibliographystyle{IEEEtran}
\bibliography{IEEEabrv,paper}

\balance

\end{document}